\documentclass[preprint,12pt]{elsarticle}

\usepackage{amssymb}
\usepackage{textcomp,booktabs}
\usepackage[usenames,dvipsnames]{color}
\usepackage{colortbl}
\definecolor{mygray}{gray}{.9}
\definecolor{mypink}{rgb}{.99,.91,.95}
\definecolor{mycyan}{cmyk}{.3,0,0,0}
\usepackage{amssymb}
\usepackage{graphicx}
\usepackage{booktabs}
\usepackage{tabularx}
\usepackage{array}
\usepackage{lscape}
\usepackage{longtable}
\usepackage{multirow}
\usepackage{amsmath}
\usepackage{colortbl}
\usepackage{epsfig}
\usepackage{epsf}

\usepackage{subfig}
\usepackage{paralist}
\usepackage{mdwlist}
\usepackage{calc}

\usepackage{algorithm}
\usepackage{algorithmicx}
\usepackage{algpseudocode}
\usepackage{multirow}
\usepackage[table,xcdraw]{xcolor}
\usepackage{eqlist}
\newcommand{\PreserveBackslash}[1]{\let\temp=\\#1\let\\=\temp}
\newcolumntype{C}[1]{>{\PreserveBackslash\centering}p{#1}}
\newcolumntype{R}[1]{>{\PreserveBackslash\raggedleft}p{#1}}
\newcolumntype{L}[1]{>{\PreserveBackslash\raggedright}p{#1}}

\journal{}

\begin{document}
\begin{frontmatter}

%% Title, authors and addresses

%% use the tnoteref command within \title for footnotes;
%% use the tnotetext command for the associated footnote;
%% use the fnref command within \author or \address for footnotes;
%% use the fntext command for the associated footnote;
%% use the corref command within \author for corresponding author footnotes;
%% use the cortext command for the associated footnote;
%% use the ead command for the email address,
%% and the form \ead[url] for the home page:
%%
%% \title{Title\tnoteref{label1}}
%% \tnotetext[label1]{}
%% \author{Name\corref{cor1}\fnref{label2}}
%% \ead{email address}
%% \ead[url]{home page}
%% \fntext[label2]{}
%% \cortext[cor1]{}
%% \address{Address\fnref{label3}}
%% \fntext[label3]{}

\title{A Physarum-inspired model for the probit-based stochastic user equilibrium problem}

%% use optional labels to link authors explicitly to addresses:
%% \author[label1,label2]{<author name>}
%% \address[label1]{<address>}
%% \address[label2]{<address>}
\author[address1]{Shuai Xu}
\author[address1]{Wen Jiang\corref{label1}}
\address[address1]{School of Electronics and Information, Northwestern Polytechnical University, Xi'an, Shannxi, 710072, China}
\cortext[label1]{Corresponding author: School of Electronics and Information, Northwestern Polytechnical University, Xi'an, Shannxi, 710072, China. Tel:+86 029 88431267; fax:+86 029 88431267. E-mail address: jiangwen@nwpu.edu.cn}

\begin{abstract}
Stochastic user equilibrium is an important issue in the traffic assignment problems, tradition models for the stochastic user equilibrium problem are designed as mathematical programming problems. In this article,  a \emph{Physarum}-inspired model for the probit-based stochastic user equilibrium problem is  proposed.
There are two main contributions of our work. On the one hand, the origin \emph{Physarum} model is modified to find the shortest path  in  traffic direction networks with the properties of  two-way traffic characteristic.  On the other hand, the modified \emph{Physarum}-inspired model could get the equilibrium flows when traveller's perceived transportation cost complies with normal distribution.
The proposed method is constituted with a two-step procedure. First, the modified  \emph{Physarum} model is applied  to get the auxiliary flows. Second, the auxiliary flows are averaged to obtain the  equilibrium flows. Numerical examples are  conducted to illustrate the performance of the proposed method, which is compared with the Method of Successive Average method.
\end{abstract}

\begin{keyword}
%% keywords here, in the form: keyword \sep keyword

%% MSC codes here, in the form: \MSC code \sep code
%% or \MSC[2008] code \sep code (2000 is the default
Traffic assignment problem, user equilibrium, elastic demand, Physarum, network
\end{keyword}

\end{frontmatter}

\section{Introduction}

The traffic assignment problem (TAP) refers to assign traffic trip of each origin-destination (OD) pair to the links in the transportation networks and give an OD trip matrix \cite{bertsekas1982projection, papageorgiou1990dynamic, yang1995traffic, ISI:000085665800003, ISI:000283918200005, du2016analysis}. Traditionally, the TAP problem falls into two major classes, known as user equilibrium(UE) and stochastic user equilibrium(SUE) \cite{sheffi1981a,smith2014the}. Considering the negative effect of road traffic congestion upon travel time, the user equilibrium was conceptualised by Wardrop \cite{Wardrop1952Some}.  Assuming that the travellers know the precise route cost and choose the route with minimum cost, the UE principle is reached when no traveller can reduce transportation cost by changing routes. To overcome the unrealistic assumption of precise perception of route travel time across travellers, the stochastic user equilibrium was firstly defined by Daganzo and Sheffi \cite{Daganzo1977On}. The SUE principle is obtained when no traveller's perceived transportation cost can be reduced  by unilaterally changing routes.

In the exist literatures, the SUE problem was classed into two types: the logit-based SUE and the probit-based SUE, according to those random costs following Gumble or normal distribution \cite{Watling2006User}. Due to explicit form and calculation, the logit-based SUE model has paid great attention \cite{Dial1971A,Fisk1980Some,Chen1991Algorithms,Torbj2007A,Liu2014Toll,Zhou2015Two} . However, the probit-based SUE model behaves more appealing, attributing to the fact that it can take no account of overlapping, or correlated routes \cite{Maher1997A}. Sheffi and Powell proposed the well-known Method of Successive Average (MSA) with
predetermined step sizes for solving the probit-based SUE problem\cite{Sheffi1982An}. Maher modified the model to decrease the computation complexity by choosing the optimal step length along the search direction\cite{Maher1997A}. Though distributed computing approaches were executed to reduce the computing time for the probit-based SUE problem \cite{Liu2013Distributed}, the development of probit-based SUE models still has some limitations, especially on large networks, which can be explained by the difficulty of completing path enumeration or Monte Carlo  simulation.

 Considering that performing the SUE principle in conventional probit-based models is difficult, we propose solving  the stochastic traffic assignment problem  by a \emph{Physarum}-inspired model. The plasmodium of \emph{Physarum} polycephalum is a large amoeboid organism, which contains a great number of nuclei and tubular structures \cite{Stephenson1995Myxomycetes}. These tubular structures will distribute protoplasm as a transportation network. Recently, it is shown that \emph{Physarum} has the capacity of finding the short path between two points in a given labyrinth \cite{nakagaki2000intelligence}. Tero \emph{et al.} \cite{Tero2010Rules} inspired an  mathematical model that can capture the basic dynamics of network adaptability through iteration of local rules and produces solutions with properties comparable to or better than those of real-world infrastructure networks. Bonifaci \cite{Bonifaci2012Physarum} has proved that the mathematical model can convergence to the shortest path. Later, the \emph{Physarum} model was used to design and simulate transport network \cite{Adamatzky2012Bioevaluation,ANDREW2013BIO,Evangelidis2015Slime,Zhang2015A}, find the short path \cite{Adamatzky2012Slime,Wang2014A,Zhang2014An,Wang2015An}.
  To handle the uncertainty in the real application \cite{Jiang2015improved},
 the \emph{Physarum} model can also solve shortest path under uncertain environment\cite{Zhangya2013fuzzy,Zhang2014fuzzy}.

 Considering the continuity of the flow and protoplasmic network adaptivity, now the UE problem can also be solved by the \emph{Physarum} model \cite{Zhang2015AnEfficient}. Note that UE problem is just a subsection of the SUE problem, here we present a  \emph{Physarum}-inspired model for the probit-based SUE problem. In the proposed model, the origin \emph{Physarum} model is modified to adapt the directed network with multiple sources and directions and the link travel time is regard as the length of \emph{Physarum}  tubular structures.

This paper is organized as follows. In Section 2, the SUE assignment problem in traffic networks is reviewed and the \emph{Physarum} polycephalum model is briefly introduced. In Section 3,  a \emph{Physarum}-inspired model for the SUE  problem is presented. In Section 4, numerical examples are given to  prove the rationality and convergence properties of the proposed model. Finally, the paper ends with conclusions in Section 5.

\section{PRELIMINARIES}\label{PRELIMINARIES}

In this section, the basic theories, including the probit-based stochastic user equilibrium problem and \emph{Physarum} polycephalum model, are briefly introduced.

\subsection{Probit-based user problem}

\subsubsection{\emph{Notations, assumptions and definitions }}

Given a strongly connected transportation network \(G = (N,A)\), where $N$ and $A$ denote the sets of nodes and links,respectively. Network attributes are denoted by notations as follow:

% \label{properties}
%\newenvironment{Description}[1]{\begin{list}{}{\renewcommand\makelabel[1]{%
%\textsl{##1}\hfil}\settowidth\labelwidth{\makelabel{#1}}\itemsep=-3pt%
%\setlength\leftmargin{%
%\labelwidth+\labelsep}}}{\end{list}}

%\begin{Description}

\begin{quote}
\begin{eqlist*}[\eqliststarinit\def\makelabel#1{#1}\labelsep1em]
\item[${R}$] Set of origin nodes, $R\subseteq N$
\item[${S}$] Set of destination nodes, $S\subseteq N$
\item[${r}$] An origin node, ${r\in R}$
\item[${s}$] An destination node, ${s\in S}$
\item[${{ K}_{rs}}$] Set of all the paths between OD pair rs.
\item[${q_{rs}}$] Travel demand between OD pair rs, and all the OD travel demands are grouped into column vector, ${\bf{q}} = {( \cdots ,{q_{rs}}, \cdots )^T}$, ${r\in R}$, ${s\in S}$
\item[\(f_k^{rs}\)] Traffic flow on path $k$ between OD pair rs, ${k\in{ K}_{rs} }$.
\item[\({\bf{f}}^{rs}\)] Column vector of traffic flows on the paths between OD pair rs, \({{\bf{f}}^{rs}} = {( \cdots ,f_k^{rs}, \cdots )^T}\),  ${k \in { K}_{rs}}$.
\item[\({\bf{f}}\)]      Column vector of traffic flows on the all paths,\({\bf{f}} = {( \cdots ,{{\bf{f}}^{rs}}, \cdots )^T}\), ${r\in R}$, ${s\in S}$.
\item[${c_k^{rs}}$]     Travel time on path $k$ between OD pair rs, ${k \in { K}_{rs}}$

\item[\({\bf{c}}^{rs}\)] Column vector of traffic time on the paths between OD pair rs, \({{\bf{c}}^{rs}} = {( \cdots ,c_k^{rs}, \cdots )^T}\),  ${k \in { K}_{rs}}$
\item[\({\bf{c}}\)]      Column vector of traffic time on the all paths,\({\bf{c}} = {( \cdots ,{{\bf{c}}^{rs}}, \cdots )^T}\), ${r\in R}$, ${s\in S}$

\item[ ${x_a}$ ]      Traffic flow on link a, \(a \in A\)
\item[ ${\bf{x}}$]    Column vector of all link flows, \({\bf{x}}{\rm{ = }}{( \cdots ,{x_a}, \cdots )^T}\), ${a\in A}$.
\item[  ${t_a}$  ]    Asymmetric travel time on link a, \(a \in A\).
\item[  ${\bf{t}}$]   Column vector of all the link-travel-time functions, \({\bf{t}}{\rm{ = }}{( \cdots ,{t_a}, \cdots )^T}\), ${a\in A}$
\item[$\delta _{a,k}^{rs}$]    $\delta _{a,k}^{rs} = 1$ if ${k\in K_{rs} }$ between OD pair rs traverses link  ${a\in A}$,  $\delta _{a,k}^{rs} = 0$, otherwise.
\item[${{\bf\triangle}^{rs}}$]    link/path incidence matrix associated with OD pair rs, ${\bf \triangle^{rs}} = (\delta _{a,k}^{rs},a \in A,k \in { K}_{rs})$
\item[${\bf\triangle}$]  link/path incidence matrix for the entire network, $ {\bf\triangle} =  {( \cdots ,{\bf\triangle}^{rs}, \cdots )} $
%\end{Description}
\end{eqlist*}
\end{quote}

According to the cost flow superposition principle, the path travel time can be valued as the summation of link travel time \cite{Sheffi1982An}, which can be expressed as:
\begin{equation}
c_k^{rs} = {t_a} \cdot \delta _{a,k}^{rs}
\end{equation}
Compactly, the relation between path travel time and link travel time can be  expressed in vector form, namely:
\begin{equation}\label{CC}
\bf{c} = {\bf\triangle}^T \cdot \bf{t}
\end{equation}

Assuming that the network users¡¯ perceived link travel time is consist of the determined link travel time and random  error term. The perceived link travel time is thus expressed as:
\begin{equation}\label{Ta}
{T_a} = {t_a}({x_a}) + {\varepsilon _a},~~~{\rm{    }}\forall a
\end{equation}
where link travel time function ${t_a}({x_a})$ is positive, continuously
differentiable and strictly monotone increasing. The error term $\varepsilon _a$ associated with link $a$ is a normally distributed random variable with zero mean for the probit-based SUE problem \cite{Sheffi1982An}, which can be expressed as following:
\begin{equation}\label{va}
{\varepsilon _a} \sim N({\rm{0,}}\gamma t_{\rm{a}}^{\rm{0}}),~~~{\rm{    }}\forall a
\end{equation}
where $\gamma$ is a proportionality constant parameter and ${t_a^0}$  is a constant which usually equals free-flow link. Similarly, the perceived link travel time is also a normally distributed random variable, namely:
\begin{equation}\label{TA}
{T_a} \sim N({\rm{t_a,}}\gamma t_{\rm{a}}^{\rm{0}}),~~~{\rm{    }}\forall a
\end{equation}
Due to the linearity of the incidence relationships, the perceived path travel time also follows a multivariate normal distribution leading to the probit model for the path choice, which can be expressed as:
\begin{equation}\label{Ck}
C_k^{rs} = \sum\limits_a^{} {{T_a}} .\delta _{a,k}^{rs},~~~{\rm{    }}\forall r,s,{k\in{ K}_{rs} }
\end{equation}

Obviously, according to the accumulation of random variables, $C_k^{rs}$ is also a normally distributed random variable with $c_k^{rs}$ mean :
\begin{equation}\label{ck}
c_k^{rs} = \sum\limits_a^{} {{t_a}} .\delta _{a,k}^{rs},~~~{\rm{    }}\forall r,s,{k\in{ K}_{rs} }
\end{equation}

 Let $f_k^{rs}$ denote the traffic flow on path $k$ between OD pair rs,  it can be expressed as the  following equation:
\begin{equation}\label{qrs}
f_k^{rs} = {q_{rs}}.P_k^{rs},~~~{\rm{    }}\forall r,s,{k\in{ K}_{rs} }
\end{equation}
where $P_k^{rs}$ denotes the path choice probability for path $k$ between OD pair rs. On the basis of economics principles, $P_k^{rs}$ denotes the probability of path $k$ being the shortest one for given path travel time between OD pair rs \cite{Qiang2012Mathematical}, namely:
\begin{equation}\label{ck}
P_k^{rs} = P(C_k^{rs} \le C_l^{rs},\forall l \ne k),~~~{\rm{    }}\forall r,s,{k\in{K}_{rs} }
\end{equation}

The stochastic user equilibrium  is reached when no user can reduce his perceived travel time by unilaterally changing routes \cite{Daganzo1977On}. The objective function of SUE problem was first proposed and proved by Sheffi and Power \cite{Sheffi1982An}:
\begin{equation}\label{Zx}
\min Z(x) =  - \sum\limits_{{\rm{rs}}} {{q_{rs}}.{\rm{E[}}\mathop {\min }\limits_{k \in {{\rm K}_{rs}}} {\rm{\{ C}}_k^{rs}{\rm{\} |}}{{\bf{c}}^{rs}}({\bf{x}}){\rm{]}}} {\rm{ + }}\sum\limits_a^{} {\{ {x_a}.{t_a}({x_a}) - \int\limits_0^a {{t_a}(w)dw} \} }
\end{equation}
where we have used the results of Williams that
\begin{equation}\label{Zx}
\frac{\partial }{{\partial {\rm{c}}_{\rm{k}}^{rs}}}E{\rm{[}}\mathop {\min }\limits_{k \in {{\rm K}_{rs}}} {\rm{\{ C}}_k^{rs}{\rm{\} |}}{{\bf{c}}^{rs}}({\bf{x}}){\rm{ ] = P}}_k^{rs}
\end{equation}

\subsubsection{\emph{The Method of Successive Average}}

The MSA algorithm developed by Sheffi and Powell \cite{Sheffi1982An} was the first algorithm applied to solve the SUE problem. In the MSA process, the link costs are calculated by the current link flows. An auxiliary link flow pattern is  produced through a stochastic network loading procedure.  And the search direction is obtained by the difference between the auxiliary link flow and the current link flow. The step size is predetermined by a descent sequence with respect to the iterations. The procedures of MSA method are summarized as following:

\begin{enumerate}[Step {1}.1: ]
\item     Choose initial link travel costs $\{ t_a^0,\forall a\}$, usually free-flow costs. Find an initial feasible flow pattern $\{ x_a^1,\forall a\} $  by carrying out, for example, a pure stochastic loading using mean costs. Set the iteration count ${n}$ to $1$.
\item     According to the current flow pattern $\{ x_a^n,\forall a\}$, calculate the current travel costs $\{ t_a^n(x_a^n),\forall a\} $.
\item  Given the mean travel costs $\{ t_a^n,\forall a\} $ and the demands of OD pairs, find the auxiliary flow pattern $\{ \hat y_a^n,\forall a\} $  by carrying out a pure stochastic loading.
\item  Calculate the new current solution according to the equation:
\begin{equation}\label{yn}
{\rm{x}}_a^{(n + 1)} = x_a^n + \frac{1}{n}(\hat y_a^n - x_a^n),~~~{\rm{    }}\forall a
\end{equation}
\item Convergence test. If the following condition is fulfilled, then stop and output  . Otherwise, ${n=n+1}$,  go to step 2.
    \begin{equation}\label{yn}
    \sqrt {\sum\limits_{a \in A} {({\rm{x}}_a^{(n + 1)}{\rm{ - }}\hat y_a^n)} }  \le {\varepsilon _{\rm{0}}}
\end{equation}
\end{enumerate}

The search direction is found by using the auxiliary flow pattern $\{ \hat y_a^n,\forall a\} $, which is computed through Monte Carlo simulation methods:
\begin{enumerate}[Step {2}.1: ]
\item Initialize counter ${i = 1}$.
\item Sample one realization from each link, using $f({T_a}|t_a^n))$.
\item Assign "all or nothing" from each origin to each destination. This results in the auxiliary flow pattern $y_a^{(i)}$.
\item  Average the flow for each link, $\bar y_a^{(i)} = [(i - 1)\bar y_a^{(i - 1)} + y_a^{(i)}]/i$.
\item  If the stopping criterion is met, set ${\rm{\hat y}}_a^n{\rm{  =   }}\bar y_a^{(i)} \forall a$; If not,set ${i=i+1}$ and go to step 2.2.
\end{enumerate}
where $f({T_a}|t_a^n)$ is the probability density of ${T_a}$, and the travel time of link ${a}$ can be calculated  according to Eq.(\ref{TA}) The stopping criterion referred to at step 2.5 may be based on the reduction of the variance of $\bar y_a^{(i)}$ as i grows  \cite{sheffi1981a,Sheffi1982An}, such as a fixed number of drawings $I_0$.

\subsection{Physarum  polycephalum model }

\emph{Physarum  polycephalum} is a single-celled amoeboid organism, which is also called as plasmodium in the vegetative phase. It is able to solve the shortest path selection, basing on its special foraging mechanism: the transformations of tubular structures and a positive feedback from flow
rates. The high rates of the flow motivate tubes to thicken, and the  diameter of the tube  minishes at a low flow rate. A total introduction for the \emph{physarum  polycephalum} model is given below.

Supposing the shape of the network formed by the Physarum represented by a graph, plasmodial tube refers to an edge of the graph and a junction between tubes refers to a node. Assuming a set of nodes ${N}$, ${N_1}$ and ${N_2}$ are signed as the source and destination nodes, any others are labeled as ${N_3}$, ${N_4}$, ${N_6}$, ${N_7}$, etc. The edge connecting nodes ${N_i}$ and ${N_j}$ is remarked as ${M_{ij}}$. The flux from node $ N_i$  to node $N_j$ through edge ${M_{ij}}$ is remarked as $Q_{ij}$, which we can expressed as   \cite{Tero2010Rules} :
\begin{equation}\label{Qij}
{Q_{ij}} = \frac{{\pi {\rm{r}}_{ij}^4}}{{8\eta {L_{ij}}}}({p_i} - {p_j}) = \frac{{{D_{ij}}}}{{{L_{ij}}}}({p_i} - {p_j})
\end{equation}
where $\eta$ is the viscosity of the fluid and ${D_{ij}}{\rm{ = }}\pi {\rm{r}}_{ij}^4/8\eta$ is measure of the conductivity of the edge $M_{ij}$
tube. $p_i$ is the measure of the pressure at the node $N_{ij}$ and $L_{ij}$ is the length of the edge of $M_{ij}$.
According to the conservation law of flow, the inflow and outflow must be
balanced, namely:
\begin{equation}\label{Q}
\sum {{Q_{ij}} = }0, ~~~{\rm{    }}({\rm{j }} \ne {\rm{1,2}})
\end{equation}
For the source nodes $N_1$ and $N_2$, the flux equations can be denoted as:
\begin{equation}\label{Q1}
\sum\limits_i {{Q_{i1}}}  + {I_0} = 0
\end{equation}
\begin{equation}\label{Q2}
\sum\limits_i {{Q_{i2}}}  - {I_0} = 0
\end{equation}
where $I_0$ is the flux from the source node to the destination node, which is assumed as a constant in the model. According to the Eqs (\ref{Qij})-(\ref{Q2}),the network Poisson equation for the pressure is derived as following:
\begin{equation}\label{Dij}
\sum\limits_i {\frac{{{D_{ij}}}}{{{L_{ij}}}}({p_i} - {p_j})}  = \left\{ \begin{array}{l}
  - 1~~~~~{\rm{  }}for~{\rm{}}j = 1, \\
  + 1~~~~~{\rm{  }}for~{\rm{}}j = 2, \\
    0~~~~~~~{\rm{    }}otherwise~~~{\rm{}} \\
 \end{array} \right.
\end{equation}
by further setting $p_2=0$ as the basic pressure level, the pressure of all nodes can be determined  according to Eq.(\ref{Dij}) and all $Q_{ij}$ can also be determined by solving Eq.(\ref{Qij}).

To accommodate the adaptive behavior of the plasmodium, the conductivity $D_{ij}$ is assumed to change when adapting to the flux $Q_{ij}$. And tubes with zero conductivity will die out. The conductivity of each tube is described as the following equation \cite{Tero2010Rules}:
\begin{equation}\label{Dt}
\frac{d}{{dt}}{D_{ij}} = f(|{Q_{ij}}|) - \alpha {D_{ij}}
\end{equation}
where $\alpha$ is the decay rate of the tube and $f$ is monotonically increasing continuous function which satisfies $f(0) = 0$. The \emph{Physarum} can converge to the shortest path when $f(|{Q_{ij}}|)=\left | Q \right |$ and $\alpha = 1$ \cite{Bonifaci2012Physarum}.
Obviously, the positive feedback exists in the model.

\section{PROPOSED METHOD}\label{Proposed method}
In this section, we employ the propoesd \emph{Physarum} model to solve the stochastic user equilibrium problem. Generally speaking, there are three problems to be addressed:
\begin{enumerate}
\item The original \emph{Physarum} model is used to solve the shortest path problem in undirected graphs (${L_{ij} = L_{ji}}$) while most network is directed graphs (${L_{ij} \ne L_{ji}}$) in real traffic assignment problem.
\item There is only one source node in the shortest path finding mode, but we should solve the traffic assignment problem with multiple sources and sinks.
\item The modified \emph{Physarum} model should approach the optimal flow distribution in the traffic assignment problem.
\end{enumerate}

\subsection{ Physarum-based model for the shortest path in the directed network }
%\begin{figure}[!ht]
%\centering
%\includegraphics[scale=0.4]{3ST.eps}
% \caption{Tubs in the directed network} \label{3STeps}
%\end{figure}

\begin{figure}[htb]

\subfloat[Origin \emph{Physarum} model]{%
  \includegraphics[width=.3\textwidth]{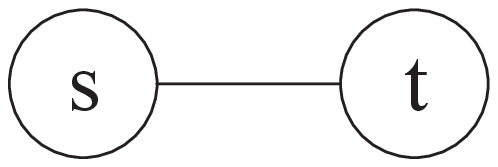}}
\subfloat[Model modified by Wang \emph{et al.}]{%
  \includegraphics[width=.3\textwidth]{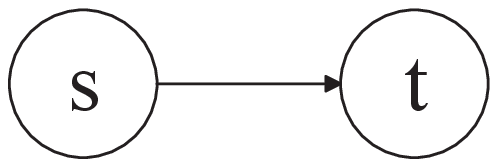}}
\subfloat[The proposed model]{%
\includegraphics[width=.3\textwidth]{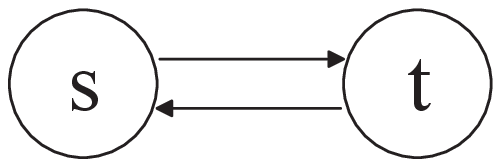}}
\caption{Three \emph{Physarum} models}
\label{3STeps}
\end{figure}

In the original \emph{Physarum} model, each arc shown in Figure \ref{3STeps}.a is bidirectional, which means the distance from node $i$ to node $j$ is same as that from node $j$ to node $i$. To solve the constrained shortest path problem, Wang \emph{et al.} \cite{Wang2014A} proposed a modified \emph{Physarum} model shown in Figure \ref{3STeps}.b where each edge is regarded as two tubes with opposite direction and equal weight. And there is only one direction between two nodes, which means that the flux can flow from node $s$ to node $t$. The modified \emph{Physarum} model has the ability to find the shortest path in the directed network. While most roads in the city have the properties of two-way traffic characteristic and opposite directions are separated with each other, the \emph{Physarum} model modified by Wang \emph{et al.} couldn't work in these networks. Here we proposed a new modified \emph{Physarum} model shown in Figure \ref{3STeps}.c. There are two opposite directions between node $s$ and node $t$, and the length of two opposite directions is denoted by ${L_{st}}$ and ${L_{ts}}$. Basing on the feature of foraging behavior, the  conductivity matrix ${D}$ implies not only the conductivity but also the direction of each tube, namely ${D_{ij} \ne D_{ji}}$ during  iterations.
In order to implement this idea into the original \emph{Physarum} model, Eq. (\ref {Dij}) is modified  as following:
\begin{equation}\label{DJI}
\sum\limits_i {(\frac{{{D_{ij}}}}{{{L_{ij}}}} + \frac{{{D_{ji}}}}{{{L_{ji}}}})({p_i} - {p_j})}  = \left\{ \begin{array}{l}
- 1~~~~~{\rm{  }}for~{\rm{}}j = 1, \\
  + 1~~~~~{\rm{  }}for~{\rm{}}j = 2, \\
    0~~~~~~~{\rm{    }}otherwise~~~{\rm{}} \\
 \end{array} \right.
\end{equation}

To keep the validity of conductivity, the conductivity equation defined in Eq.(\ref{Qij}) should be improved as following:
\begin{equation}\label{QIJ}
{Q_{ij}} = \left\{ \begin{array}{l}
 \frac{{{D_{ij}}}}{{{L_{ij}}}}({p_i} - {p_j}),~~~~~{\rm{  }} \frac{{{D_{ij}}}}{{{L_{ij}}}}({p_i} - {p_j}) > 0 \\
0,~~~~~~~~~~~~~~~~~~~~~~~~{\rm{    }}otherwise~~~{\rm{}} \\
 \end{array} \right.
\end{equation}
Particularlly when ${L_{st}} = inf$  or ${L_{ts}} = inf$, which means that the flux can only flow from node $t$ to node $s$ or from node $s$ to node $d$, our modified model is the same as that of Wang \emph{et al.} Exactly, the model modified by Wang \emph{et al.} is a section of our modified model.

\subsection{Physarum-based model for multiple sources and directions }

In the original \emph{Physarum} model, there is only one source node and one direction node. While in the stochastic user equilibrium problem, there are always multiple OD pairs. Assuming ${O}$ denoting the set of origin nodes, $O \subseteq N$, and ${D}$ denoting the set of destination nodes, $D \subseteq N$,  we can modify Eqs. (\ref {Q1}) and (\ref{Q2}) as following:
\begin{equation}\label{Qio}
\sum\limits_i {{Q_{io}}}  + {I_o} = 0,~~~{\rm{    }}o \in O
\end{equation}
\begin{equation}\label{Qid}
\sum\limits_i {{Q_{id}}}  - {I_d} = 0,~~~{\rm{    }}d \in D
\end{equation}
where ${I_o}$ is the inflow at the origin node ${o}$,${I_d}$ is the outflow at the destination node ${d}$. To ensure the flow is distributed in a optimal way, here we use the modified model proposed by Zhang \cite{Zhang2015A} to replace Eq.(\ref{Dij}):
\begin{equation}\label{DIJ}
\sum\limits_i {(\frac{{{D_{ij}}}}{{{L_{ij}}}} + \frac{{{D_{ji}}}}{{{L_{ji}}}})({p_i} - {p_j})}  = \left\{ \begin{array}{l}
  - {I_o},~~~~~{\rm{  }}\forall o \in O, \\
  + {I_d},~~~~~{\rm{  }}\forall d \in D, \\
 0,~~~~~~~~~{\rm{  }}otherwise \\
 \end{array} \right.
 \end{equation}

\subsection{ Physarum-inspired model for the probit-based SUE problem }

Now, we study how to solve the probit-based stochastic user equilibrium problem basing the \emph{Physarum}-inspired model. Due to the feature of foraging behavior, the flow and the conductivity along each link are continuous in the process of \emph{Phyasrun} approaching the shortest path. While in other classical shortest path algorithms, such as Dijkstra algorithm \cite{Dijkstra1959A}, Floyd algorithm \cite{Floyd1962Algorithm}, algorithms approach the shortest path by traversing all the nodes until the destination node is visited, which is totally uncontinuous.

Considering the continuity  and dynamic reconfiguration of \emph{Physarum} model, we can update the link travel time within each  iteration. The flux will be redistributed by the modified \emph{Physarum} model when the link travel time is updated during iterations. Here we adopt the following equation to update the length of link $a$:
\begin{equation}
C_a^n = \frac{{C_a^{n - 1} + {t_a}({x_a})}}{{\rm{2}}}
\end{equation}

where $x_a$ denotes the traffic flow on link $a$ at the $(n)_{th}$ iteration, $C_a^{n}$ and $C_a^{n-1}$ represent the length of link $a$ at the $n_{th}$ and $(n-1)_{th}$ iteration.
And the search direction of link length  $C_a$ is guided by ${t_a}(x_a)$. Note that in equilibrium, there will be $C_a=t_a(x_a)$, which means the length of link $a$ equals the travel time along link $a$.

The main steps of the proposed method for the probit-based stochastic user equilibrium problem is presented in Algorithm 1.
In the process of Monte Carlo simulation, we can use \emph{Physarum} model to replace "All or nothing" method to calculate auxiliary flow pattern. In the iteration, $C^n$ is the current link travel cost matrix at the $n$ iteration, $C_{ij}^n$ represents the current travel cost from node $i$ to node $j$, at the $n$ iteration. Particularly, $C^0$ is the free-flow link travel cost matrix, $C_{ij}^0$ represents the free-flow travel cost from node $i$ to node $j$.

Different from the MSA algorithm, the current travel cost($C_{ij}^n$) is calculated by the modified auxiliary flow in the proposed algorithm, attributed to the continuity of  \emph{Physarum} model. During the the process of Monte Carlo simulation£¬ the flow and the conductivity along each link are continuous.

 More importantly, the current travel cost $C_{ij}^n$ doesn't equal  $C_{ji}^n$ when $\hat Q_{ij}^{(I_0)} \ne \hat Q_{ji}^{(I_0)}$, which means the same edge have different travel costs in two opposite directions. This peculiarity is rather important in nowadays traffic network. Because most roads in the city don't interfere  in two opposite directions, opposite directions are separated with each other. Indeed, the flow in edge $L_{ij}$ does't influence the travel cost in edge $L_{ji}$.

\begin{algorithm}
  \caption{a Physarum-inspired model for the probit-based SUE problem}
  \begin{algorithmic}
   \State // $\varepsilon_{0}$ is the stopping criterion of the whole method.
   \State // $I_o$ is the stopping criterion of the  Monte Carlo simulation,also called as the inner iteration.
   \State // $n$ is called as the outer iteration.
   \State // $\bar Q^{n}$ is the current flow matrix at the $n$ iteration, $\bar Q_{ij}^{n}$ represents the current flow from node $i$ to node $j$.
   \State // $\hat Q^{(I_0)}$ is the modified auxiliary flow matrix.
   \State $D_{ij} = [0.5,1](\forall i,j = 1,2,\cdots,N \wedge C_{ij}^0 \ne 0)$
   \If {$C_{ij}^0 == inf $}
        \State $D_{ij} = 0$
   \EndIf
   \State $Q_{ij} = 0(\forall i,j = 1,2,\cdots,N )$
    \State ${\bar Q_{ij}^0} = 0(\forall i,j = 1,2,\cdots,N )$
   \State $p_{ij} = 0(\forall i,j = 1,2,\cdots,N )$
   \State $n=1$ //Iteration counter
   \While{$\varepsilon \le \varepsilon_0$}
        \State  $C_{ij}^n= \frac{C_{ij}^{n-1}+ t_{ij}(\hat Q_{ij}^{(I_0)})}{2}(\forall i,j = 1,2,\cdots,N )$
        \State $i=1$ //Monte Carlo simulation counter
            \While{$i \le I_0$}
              \State  $L_{ij}=N(C_{ij}^n,\gamma C_{ij}^0 )~~~(\forall a \in A)$~~~//Using Eq.(\ref{TA})
              \State //Calculate the pressure of every node using Eq.(\ref{DIJ})
              \State
              $\sum\limits_i {(\frac{{{D_{ij}}}}{{{L_{ij}}}} + \frac{{{D_{ji}}}}{{{L_{ji}}}})({p_i} - {p_j})}  = \left\{ \begin{array}{l}
          - {I_o},~~~~~{\rm{  }}\forall o \in O, \\
          + {I_d},~~~~~{\rm{  }}\forall d \in D, \\
         0,~~~~~~~~~{\rm{  }}otherwise \\
         \end{array} \right.$
             \State //Calculate the flux of every edge using Eq.(\ref{QIJ})
             \State
             ${Q_{ij}} = \left\{ \begin{array}{l}
         \frac{{{D_{ij}}}}{{{L_{ij}}}}({p_i} - {p_j}),~~~~~{\rm{  }} \frac{{{D_{ij}}}}{{{L_{ij}}}}({p_i} - {p_j}) > 0 \\
        0,~~~~~~~~~~~~~~~~~~~{\rm{    }}otherwise~~~{\rm{}} \\
         \end{array} \right.$
            \State ${\rm{D}}_{ij}^{i + 1} = ({\rm{D}}_{ij}^i + {Q_{ij}})/2$
             \State $\hat Q^{(i)} = [(i-1)\hat Q^{(i-1)} + Q^{(i)} ]/i$
              \State $i=i+1$
        \EndWhile
     \State $\bar Q^{n} = [(n-1)\bar Q^{n-1} + \hat Q^{(I_0)} ]/n$
       \State ${\varepsilon }=\sqrt {\sum\limits_{i,j\in N} (\bar Q_{ij}^{n}{\rm{ - }}\bar Q_{ij}^{n-1} )}$
       \State $n = n +1$
   \EndWhile
  \end{algorithmic}
\end{algorithm}

\section{NUMERICAL EXAMPLES}\label{numerical examples}

In this section, two examples are designed to prove the rationality and convergence properties of the proposed algorithm, a one source and sink node network and a multiple sources and sinks network. The inner iteration $I_0$ and the outer iteration $n$ are compared with those in the MSA algorithm.

\begin{figure}[!ht]
\centering
\includegraphics[scale=0.4]{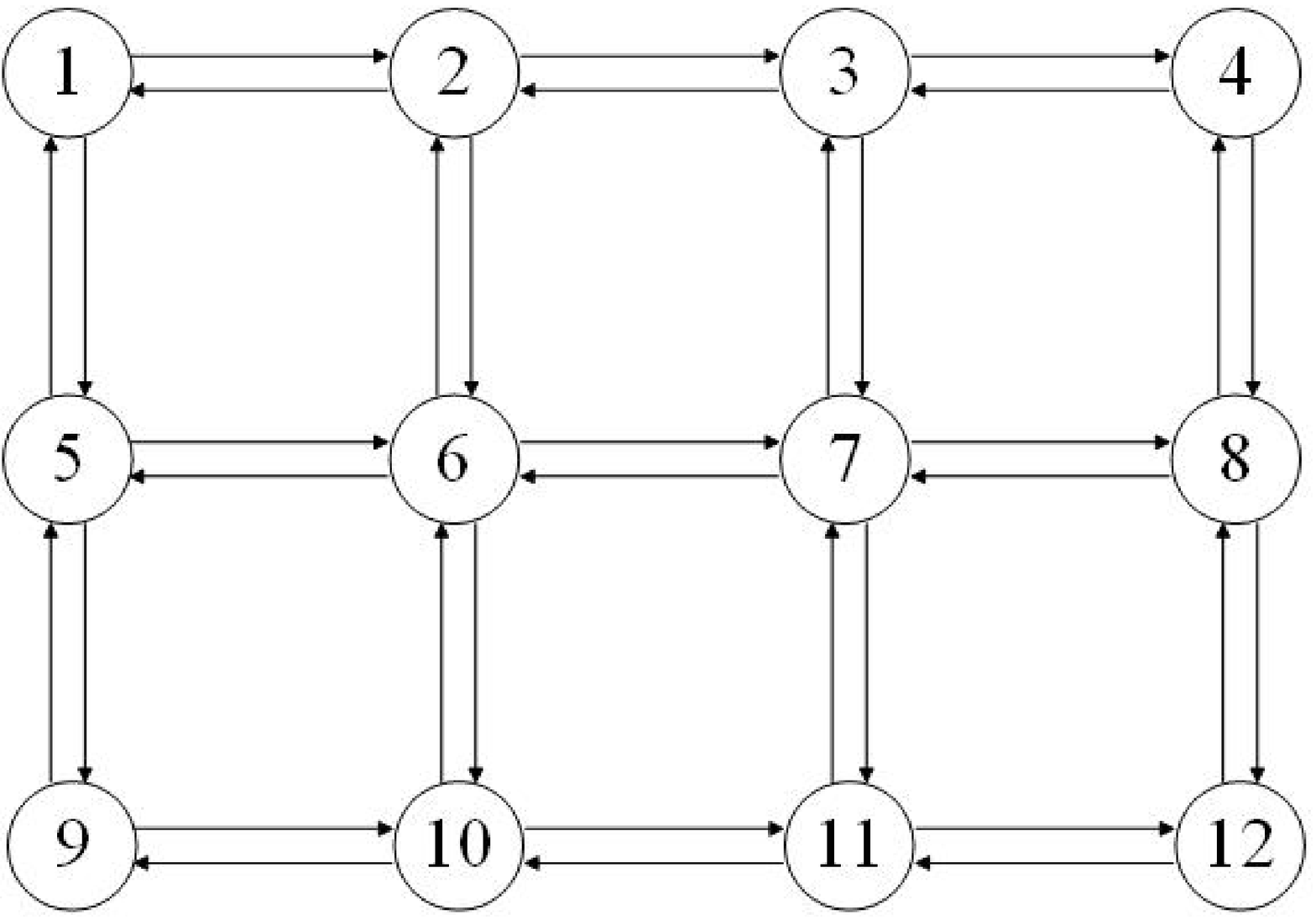}
 \caption{A simple network with 12 nodes used in  \cite{Sheffi1982An}} \label{FIG1}
\end{figure}

Both tests are investigated using a simple network shown in Figure \ref{FIG1}, which is introduced by Sheffi and Powell \cite{Sheffi1982An}.  Link costs are calculated  by the US Bureau of Public Roads (BPR) function, which is
expressed as following:
\begin{equation}\label{taxa}
{{\rm{t}}_{\rm{a}}}({x_a}) = {\alpha _a} + {\beta _a}{x_a}^4
\end{equation}
where parameters of the link cost functions for each link, $\alpha_a$ and$\beta _a$ are shown in Table \ref{Table1}.

\begin{table}[!h]
{\footnotesize
\caption{Parameters of the link cost functions for each link, ${{\rm{t}}_{\rm{a}}}({x_a}) = {\alpha _a} + {\beta _a}{x_a}^4$}\label{Table1}
\begin{tabular*}{\columnwidth}{@{\extracolsep{\fill}}@{~~}llll@{~~}}
\toprule
  From node & TO node & $\alpha _a$ & $\beta _a$\\
\midrule
  1         & 2      & 20     & 0.0056    \\
  1         & 5      & 18     & 0.0078     \\
  2         & 1      & 20     & 0.0071     \\
  2         & 6      & 19     & 0.0033    \\
  2         & 3      & 23     & 0.0086   \\
  3         &2       &23       &0.0108\\
  3         &7       &16         &0.0101\\
  3         &4      &17         &0.0063\\
  4         &3      &17         &0.0116\\
  4         &8      &22         &0.0138\\
  5         &1      &18         &0.0131\\
  5         &6      &14         &0.0093\\
  5         &9      &24         &0.0026\\
  6         &2      &19         &0.0048\\
  6         &5      &14         &0.0041\\
  6         &7      &17         &0.0123\\
  6     &10     &20     &0.0056\\
  7     &3      &16     &0.0078\\
  7     &6      &17     &0.0071\\
  7     &8      &13     &0.0033\\
  7     &11     &26     &0.0086\\
  8     &4      &22     &0.0108\\
  8     &7      &13     &0.0101\\
  8     &12     &19     &0.0063\\
  9     &5      &24     &0.0016\\
  9     &10     &7      &0.0138\\
  10    &9      &7      &0.0131\\
  10    &6      &20     &0.0093\\
  10    &11     &18     &0.0026\\
  11    &10     &18     &0.0048\\
  11    &7      &26     &0.0141\\
  11    &12     &17     &0.0123\\
  12    &8      &19     &0.0056\\
  12    &11     &17     &0.0078\\
\bottomrule
\end{tabular*}
}
\end{table}

 According to Eq.(\ref{TA}), the value of $\gamma$ determines the variance of the perceived link travel cost ,which has a great effect on the convergence properties of the algorithm. In both tests, $\gamma$ is kept constant at 0.3. In Eq.(\ref {yn}), $\varepsilon _0$ is the condition of stopping iteration, if the value of $\varepsilon _0$ is to small, the process of both algorithms cost much calculating time. However, we can't get the final travel flux if its value is too large.  Hence, in order to compare the speed of convergence, the value of $\varepsilon _0$ is kept constant at 0.1 in the proposed algorithm and the MSA algorithm. And all computational experiments are executed using Matlab on Intel(R) Core(TM) i5-5200U processor (2.2Ghz) with 8.00 GB of RAM under Win 8.

\subsection{Example 1} \label{Example1}

\begin{figure}[!ht]
\centering
\includegraphics[scale=0.4]{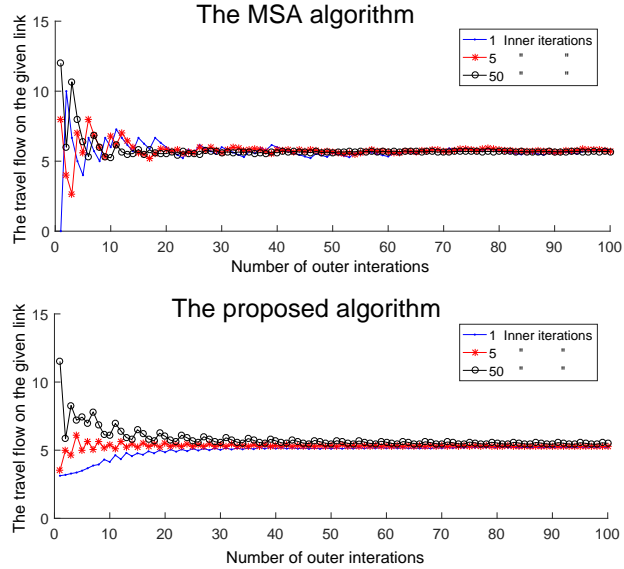}
 \caption{Convergence of the proposed algorithm vs the MSA algorithm } \label{FIGfinal}
\end{figure}

In this example, there is only one source-direction from node 1 to node 12 with travel rate of $q_{1,12}= 20$ vehicles per unit time.  To study the effect of the inner iteration $I_0$, we examined the flow on particular link   corresponding the number of inner iterations. Here, we choose the traffic flow of link $L_{6,7}$  in both algorithms. The effect of  inner iterations is illustrated in Figure \ref{FIGfinal}.

It's clearly that the convergence per equilibrium iteration improves when the  inner iteration $I_0$ augments. However, in the proposed algorithm, the deviation between lower iteration and equilibrium flow is obviously smaller than that in the MSA algorithm when the counter of outer iterations is small.

What's importantly, both algorithms can get the same equilibrium flow when the inner iterations are different. Sheffi and Powell has proved that very few inner simulation iterations (possibly just one) may be sufficient to achieve a reasonable convergence rate of  the equilibrium iterations by the SMA algorithm \cite{Sheffi1982An}. Note that the proposed algorithm can also get the similar result when the inner iteration equals 1 in Figure \ref{FIGfinal}, we speculate that one iteration can also achieve a reasonable convergence rate of  the equilibrium iterations. The numerical example is presented as below.

To  prove the rationality and convergence properties of the proposed algorithm, the inner iteration $L_0$ should be kept same in both algorithms. So we should keep the inner iteration $I_0$ equal 1 in both algorithm. The results of link traffic flow calculated by both algorithms are shown in Table \ref{Table2}.
\begin{table}[!h]
{\footnotesize
\caption{The link traffic flow  calculated by both algorithms in
\ref{Example1}}\label{Table2}
\begin{tabular*}{\columnwidth}{@{\extracolsep{\fill}}@{~~}llllllll@{~~}}
\toprule
  From node & TO node & The MSA algorithm &The proposed algorithm\\
\midrule
  outer iteration $n$  &$~~~$  &12233  &236      \\
\midrule
  computer time(s)   &    & 2.317002 &0.069317    \\
  \midrule
  1         & 2      & 10.3639     & 10.2070    \\
  1         & 5      & 9.6361     & 9.5445    \\
  2         & 1      & 0         & 0    \\
  2         & 6      & 4.4459    & 4.4894  \\
  2         & 3      & 5.9180     &5.7079   \\
  3         &2       &0      &0\\
  3         &7       &2.7803        &2.5665\\
  3         &4      &3.1377         &3.1324\\
  4         &3      &0         &0\\
  4         &8      &3.1377        &3.1328\\
  5         &1      &0         &0\\
  5         &6      &4.9213         &4.7524\\
  5         &9      &4.7148         &4.7896\\
  6         &2      &0              &0\\
  6         &5      &0              &0\\
  6         &7      &5.6918         &5.4874\\
  6     &10     &3.6754     &3.7607\\
  7     &3      &0     &0\\
  7     &6      &0     &0\\
  7     &8      &7.6230    &7.5404\\
  7     &11     &0.8492     &0.5210\\
  8     &4      &0     &0\\
  8     &7      &0     &0\\
  8     &12     &10.7607     &10.6752\\
  9     &5      &0     &0\\
  9     &10     &4.7148      &4.7948\\
  10    &9      &0      &0\\
  10    &6      &0     &0\\
  10    &11     &8.3902     &8.5612\\
  11    &10     &0     &0\\
  11    &7      &0     &0\\
  11    &12     &9.2393     &9.0669   \\
  12    &8      &0     &0\\
  12    &11     &0     &0\\
\bottomrule
\end{tabular*}
}
\end{table}

It's obviously there are same traffic paths in the network. And traffic rates in each link are almost similar calculated by both algorithms, which differ by no more than $0.32$  vehicles per unit time. Besides, the convergence rate of the proposed algorithm is faster than that of the SMA algorithm. This peculiarity will become much more important especially when the network is quite large. Considering that responsiveness of the traffic flow assignment is much more significant in nowadays traffic network, the proposed algorithm contributes a positive idea to reduce computing time.

\subsection{Example 2}\label{Example2}

In this example, we also used the traffic network shown in Figure \ref{FIG1} and Table \ref{Table1}. Different from Example 1, the origin-destination demands are assumed as $q_{1,12}= 10$ and $q_{1,8}= 10$, which denote the rate of vehicles per unit time. The inner iteration  was set as $10$ in this example. The results of link traffic flow calculated by both algorithms are shown in Table \ref{Table3}.
Clearly, traffic rates calculated by both algorithms  are also similar  and the proposed method obviously get the equilibrium flow faster than the MSA algorithm. The maximum error of the link flows is no more than $0.35$  vehicles per unit time. The computing time don't increase compared with that in Example 1.

\begin{table}[!h]
{\footnotesize
\caption{The link traffic flow  calculated by both algorithms in \ref{Example2} }\label{Table3}
\begin{tabular*}{\columnwidth}{@{\extracolsep{\fill}}@{~~}llll@{~~}}
\toprule
  From node & TO node & The MSA algorithm &The proposed algorithm\\
\midrule
  outer iteration $n$  &$~~~$  &686  &179      \\
\midrule
  computer time(s)   &    &2.162422 &0.068314      \\
  \midrule
  1         & 2      & 10.3988     & 10.1945    \\
  1         & 5      & 9.6058    & 9.4830   \\
  2         & 1      & 0         & 0    \\
  2         & 6      & 3.6292    & 3.5431 \\
  2         & 3      & 6.7686     &6.6450   \\
  3         &2       &0      &0\\
  3         &7       &2.2849      &2.1953\\
  3         &4      &4.4803        &4.4454\\
  4         &3      &0         &0\\
  4         &8      &4.4803       &4.4424\\
  5         &1      &0         &0\\
  5         &6      &5.1153        &4.7598\\
  5         &9      &4.4905        &4.7273\\
  6         &2      &0              &0\\
  6         &5      &0              &0\\
  6         &7      &6.4263        &6.3017\\
  6     &10     &2.3182     &1.9937\\
  7     &3      &0     &0\\
  7     &6      &0     &0\\
  7     &8      &8.7109    &8.4797\\
  7     &11     &0     &0.0325\\
  8     &4      &0     &0\\
  8     &7      &0     &0\\
  8     &12     &3.2044     &3.0647\\
  9     &5      &0     &0\\
  9     &10     &4.4905      &4.7273\\
  10    &9      &0      &0\\
  10    &6      &0     &0\\
  10    &11     &6.8088     &6.7691\\
  11    &10     &0     &0\\
  11    &7      &0     &0\\
  11    &12     &6.8088     &6.7691  \\
  12    &8      &0     &0\\
  12    &11     &0     &0\\
\bottomrule
\end{tabular*}
}
\end{table}

\section{CONCLUSIONS}\label{conclusions}

Considering of deviation between traveller's perceived transportation cost and actual cost, the stochastic user equilibrium is much more significant than user equilibrium. This paper  presents a \emph{Physarum}-inspired model for the probit-based stochastic user equilibrium problem. The \emph{Physarum} model is modified to solve the SUE problem in the first time. To satisfy the characteristic of the real traffic networks, the origin \emph{Physarum} model is modified to find the shortest path in direction networks with multiple sources and directions. Considering  the foraging behavior of \emph{Physarum}, the \emph{Physarum} could find the shortest travel time path between each OD pair. The equilibrium flows could be obtained when\emph{Physarum} couldn't find a shorter travel time path.

We compared the proposed algorithm with the MSA algorithm. And Numerical results showed that the proposed algorithm can effectively achieve the SUE solution in practice. If the inner iteration properly  assigned, the proposed algorithm is faster and more efficient than the MSA algorithm. Note that many investigations about paralleled \emph{Physarum} model have been achieved \cite{adamatzky2008towards}, the time consumption of the proposed algorithm will obviously reduced in concurrent computation.
Besides, the proposed  method is easy to combine with other algorithms \cite{ISI:000372944400002}.

%% The Appendices part is started with the command \appendix;
%% appendix sections are then done as normal sections
%% \appendix

%% \section{}
%% \label{}

%% References
%%
%% Following citation commands can be used in the body text:
%% Usage of \citep is as follows:
%%   \citep{key}          ==>>  [#]
%%   \citep[chap. 2]{key} ==>>  [#, chap. 2]
%%   \citept{key}         ==>>  Author [#]

%% Refeyjgukkurences with bibTeX database:

\section*{ACKNOWLEDGMENTS}

The work is partially supported by National Natural Science Foundation of China (Grant No. 61671384), Natural Science Basic Research Plan in Shaanxi Province of China (Program No. 2016JM6018), the Fund of SAST (Program No. SAST2016083), the Seed Foundation of Innovation and Creation for Graduate Students in Northwestern Polytechnical University (Program No. Z2016122).

\bibliographystyle{model1-num-names}
\bibliography{SxxSxx}
\end{document}